\documentclass[seceq]{ptptex}
\usepackage{relsize}
\usepackage{amsmath}
\usepackage{amsfonts}
\usepackage{graphicx}

\newcommand{\xscrpt}[1]{{\textstyle \mbox{\scriptsize #1}}}

\title{
Drawbacks of applying perturbative schemes to meson spectroscopy \footnote{Based on notes for a talk
presented by Eef van Beveren at the {\it 11th International Conference
on Meson-Nucleon Physics and the Structure of the Nucleon},
September 10-14 (2007) IKP Forschungzentrum J\"{u}lich (Germany).}%
}

\author{
K. P. \textsc{Khemchandani}$^{1,}$\footnote{ Present address: Research Center
for Nuclear Physics (RCNP), Ibaraki, Osaka 567-0047, Japan.},
 Eef van \textsc{Beveren}$^{1}$, George \textsc{Rupp}$^{2}$%
}

\inst{
$^{1}$Centro de F\'{\i}sica Computacional,
Departamento de F\'{\i}sica, Universidade de Coimbra, P-3004-516 Coimbra, Portugal\\
$^{2}$Centro de F\'{\i}sica das Interac\c{c}\~{o}es Fundamentais,
Instituto Superior T\'{e}cnico, Universidade T\'{e}cnica de Lisboa, 
P-1049-001 Lisboa, Portugal
}

\abst{
We study meson-meson scattering in a soluble model which describes
asymptotically free mesons and confined quark-antiquark pairs via coupled
channels. Concretely, the two scattered mesons are assumed to interact
through $s$-channel meson-exchange diagrams.
Furthermore, we develop a perturbative expansion of the model,
and show that the thus found resonance pole positions, including
contributions up to fourth order in perturbation theory, completely
fail to reproduce the exact results.
This shows that the resonance predictions based on perturbative
approximations in quark models may be highly unreliable.
}

\begin{document}

\maketitle

\section{Introduction}
Hadron spectroscopy in present times appears to have become subdivided
into many different fields of interest, like conventional quark
spectroscopy, hadronic molecular states, dynamically generated resonances,
tetraquarks and pentaquarks, glueballs, gluonic hybrids, and so forth.
The advancement of detector and analysis techniques at the many new
experimental facilities in the intermediate-energy range has been resulting
in the observation of more and more hadronic states,
many of which no not seem to fit into the traditional quark spectrum of
$q\bar{q}$ mesons and $qqq$ baryons. This has led to the investigation of
other possible configurations that might be viable within QCD.
On the one hand, this makes it important and interesting
to explore the structure and properties of hadrons, which may shed light on
the dynamics of strong interactions at low energies.
On the other hand, though, a very careful analysis is required, in order to
avoid confusions and controversies.
We are convinced that the most important need of present-day research in
hadron spectroscopy is some consensus on the ideal quark model to confront
with the data. And right now is probably the best moment
to tackle this problem, since quite a lot of data,
at least concerning meson spectroscopy, is being produced
(see for example
Refs.~\cite{NPA827p291C,STosi2,ARXIV08103829,
AIPCP1182p455,ARXIV09103404,LATHUILE02p569,
NPPS186p371,ARXIV10012252,ARXIV09065333}),
and even more data, of better quality, is expected in the near future.
All this could help gather a vastly improved understanding of quarkonia
and other mesonic resonances.
However, in such a situation it is extremely important
to first agree on the hadron spectrum, as obtained from a chosen quark
model, since only then valid conclusions can be drawn about
possible incompatibilites with standard quark configurations.

Indeed a lot of work is being done on trying to understand
the conventional and unconventional structure of different
``exotic'' hadrons. For example, a very interesting analysis of the
quark content and possible molecular nature of many newly found mesonic
states has been carried out on the basis of QCD sum rules
\cite{PRD80p056002,ARXIV09111958}, while
an attempt to distinguish between quarkonium and hybrid states
was made in Ref.~\cite{PLB657p49}.
Another very interesting investigation, namely of the quark
and molecular content of baryon resonances,
was reported in Ref.~\cite{PRC78p025203}.
Furthermore, Ref.~\cite{PRD80p074028} nicely explained
the phenomenon of dynamically generated resonances
and the concept of dynamical reconstruction,
in order to be able to pinpoint (dominantly) $q\bar{q}$ states.
All these meticulous efforts and the corresponding results 
would get more merit, if a universally agreed
conventional hadron spectrum was known.

The main issue is that experimental evidence for possible resonances
is obtained from total or partial-wave cross sections,
as well as angular distributions and decay modes.
In order to interpret the data, one needs a model, since perturbative
QCD cannot be used  at low energies.
In the present paper, we will focus on
non-exotic meson-meson scattering. Nevertheless, the results can easily
be generalized for application to hadron-pair production \cite{AP323p1215}.
We intend to describe the cross sections and resonance pole positions
for meson-meson scattering in an as large as possible energy range,
rather than focusing on just one peak.
At this point enters the main philosophy behind the model,
namely that the enhancement structure of
cross sections in non-exotic meson-meson scattering
stems from the quark-antiquark spectrum.
Consequently, for such reactions we must study a coupled system
of a $q\bar{q}$ state and non-exotic two-meson channels.
This is an absolutely minimal requirement for modeling
the cross sections in this case.
Further extensions to multiquark or hybrid resonances might be contemplated
in case the minimal model turns out not to reproduce the experimental data
sufficiently well.
Furthermore, the proposed strategy \cite{PRD21p772,PRD27p1527}
simultaneously covers two-meson molecules and $q\bar{q}$ systems.
Hence, the physical solutions are not
just pure $q\bar{q}$ or molecular states,
but rather mixtures of these two configurations.
One could then try to find out which component is dominant
\cite{ZPC19p275,PRD44p2803,ARXIV10073461}.
However, that lies beyond the scope of this paper.
In more popular terms, one might refer to
the states obtained from such a model
as $q\bar{q}$ systems surrounded by a meson cloud.
In the past, our approach was also called the unitarization scheme
for the quark-antiquark system
\cite{Cargese75p305,AP123p1,PRL49p624,PRD29p110}.

The model we are going to use treats confined quarks and hadronic
decay channels on an equal footing, via coupled channels, regardless
of whether the energy is above or below the thresholds of the decay channels.
It was developed in Refs.~\cite{PRD21p772,
PRD27p1527,ZPC21p291,EPJC22p493,AP324p1620},
and has been extensively used to study the properties of mesonic resonances
(for some of the recent works, see
Refs.~\cite{EPL85p61002,ARXIV10052486,
ARXIV10052490,PRD80p094011,ARXIV08121527,PRD80p074001}
and references therein).
The effective meson-meson potential in the model
consists of $s$-channel exchange
of a confined $q\bar{q}$ pair, with radial quantum
number running from 0 to infinity and orbital angular momentum
compatible with total $J^{PC}$.
The importance of hadron dynamics in understanding the meson 
spectrum for low and intermediate energies already becomes obvious from the fact
that it easily gets energetically favorable for the ``string'' between
the quark and the antiquark to break, alongside the creation of a new and light
$q\bar{q}$ pair, which then may lead to hadronic decay. An even stronger
indication comes from the light scalar mesons, whose unconventional nature
is related to their very strong coupling to $S$-wave two-meson channels
\cite{ZPC30p615}. The latter work also shows that the inclusion of 
bare-meson exchange in the $s$-channel, in addition to the meson-meson contact
interaction
\cite{PRL49p624,PRD60p074023,PRD59p074001,ARXIV08050552,ARXIV08054803},
in certain cases leads to finding new states which might be absent when
considering contact interactions only.
This phenomenon was explained very neatly
in Ref.~\cite{AP324p1620},
where it was shown that, for the study of a restricted energy range
corresponding to a particular resonance,
the contribution from different diagrams
involving meson exchange with different quantum numbers
gives rise to a constant interaction, which is equivalent
to considering a contact interaction in unitarized models.
It was further shown in Ref.~\cite{AP324p1620} that, in order to
understand a larger energy range, covering several resonances,
 meson-exchange diagrams are required as well.
This explains why the common use of contact interactions in unitarized models,
to study dynamically generated hadron resonances, works quite well
\cite{PRD80p114013,PRD76p074016,ARXIV10050283,PRC77p042203,
EPJA37p233,ARXIV10030364,PRD80p094012,PRC79p065207,PRC80p055206}.
However, from Ref.~\cite{AP324p1620} it becomes clear
that the development of a broader and more general perspective
for hadron spectroscopy requires the
treatment of quarks and hadrons as coupled systems.
In this paper, we will show that not only the handling
of coupled mesons and quarks is necessary,
but also the full solution of the scattering equations is essential.
In particular, we will demonstrate that approximating resonance pole
positions perturbatively leads to unreliable results.

In the next section, we will first describe the exact formalism,
followed by the construction of a perturbative expansion thereof.
In the subsequent sections, we will choose some specific examples
to show that no meaningful results can be obtained 
from the perturbative expansion.
Finally, we will summarize the detailed discussions
in the paper.

\section{Formalism}
\label{Formalism}
We study meson-meson scattering using a model
in which quarks and hadrons are considered coupled systems.
The formalism amounts to solving a scattering equation for mesons,
with the lowest-order term of the Born series given by an
effective interaction due to the exchange of a confined $q\bar{q}$ pair.
The potential between the latter pair is written
in terms of a harmonic oscillator. The eigenenergies of the harmonic
oscillator thus correspond to the bare $q\bar{q}$ spectrum.
The effective meson-meson potential, which is the lowest-order term of the
full scattering amplitude, involves an infinite sum over all the diagrams
with $s$-channel $q\bar{q}$ exchange, having different radial quantum number $n$
($0\leq n < \infty$), as shown in Fig.~\ref{fig1} below.
\begin{figure}[htbp]
\centering
\includegraphics[scale=0.45]{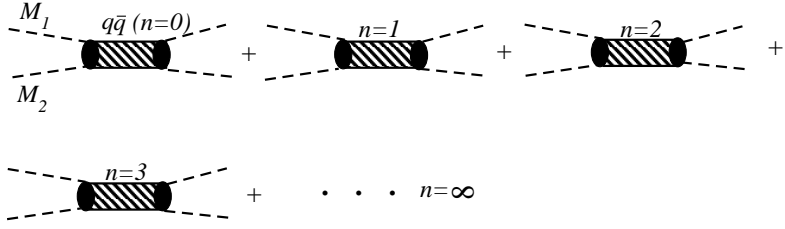}
\caption[]{The meson-meson potential,
i.e., the Born term of the full scattering amplitude,
which involves the exchange of a confined $q\bar{q}$ pair, 
with radial quantum number $n$
running from 0 to $\infty$.}
\label{fig1}
\end{figure}

Although not strictly necessary, it is illustrative
to consider a formulation of the model
in terms of a coupled system of nonrelativistic Hamiltonians.
However, a rigorous derivation in terms of a sum of meson loops,
which leads to the same final result, is also possible.

Since the model is based on coupling a confined quark-antiquark pair
and a meson pair, we describe the system by the equations
\begin{eqnarray}
H_{c}\psi_{c}(\vec{r})+V_{T}(\vec{r})\psi_{f}(\vec{r}) &=&
E\psi_{c}(\vec{r}),
\label{hc}\\
H_{f}\psi_{f}(\vec{r})+V_{T}(\vec{r})\psi_{c}(\vec{r}) &=&
E\psi_{f}(\vec{r}),
\label{hf}
\end{eqnarray}
where the subscripts ``$c$'' and ``$f$'' (here and throughout this article)
refer to the confined quarks and free mesons
(i.e., considering them plane waves), respectively,
and $V_{T}$ is the transition potential between the two sectors.

$H_{c}$ and $H_{f}$ describe the Hamiltonians of these sectors, reading
\begin{eqnarray}
H_{c} &=&
-\frac{\bigtriangledown_{r}^{2}}{2\mu_{c}}+m_{q}+m_{\bar{q}}+V_{c}(r),\\
H_{f} &=&
-\frac{\bigtriangledown_{r}^{2}}{2\mu_{f}}+M_{1}+M_{2},
\end{eqnarray}
where the confining potential is assumed to be a harmonic oscillator, viz.\
\begin{equation}
V_{c}=\frac{1}{2}\mu_{c}\omega^{2}r^{2},
\end{equation}
with $\mu_{c}$ and $\omega$ the reduced mass and frequency
of the $q\bar{q}$ system, respectively.
Furthermore, $M_{1}$, $M_{2}$, and $m_{q}$, $m_{\bar{q}}$
are the meson and quark masses, respectively.
Our choice of a harmonic-oscillator potential in the confined sector
is based on earlier observations of regular spacings 
in the quarkonium spectra \cite{PRD21p772,PRD27p1527}, which seem to be
confirmed by states found in even the most recent experiments
\cite{EPL85p61002,ARXIV09044351,ARXIV09062278,ARXIV10053490}.
In writing down the equations above, we have assumed only one $q\bar{q}$
and one meson-meson channel, for the sake of simplicity.
These equations can be straighforwardly generalized to the
multichannel case, taking then a matrix form \cite{PRD80p094011}.

Now, Eqs.~(\ref{hf}) and (\ref{hc}) can be rewritten as
\begin{eqnarray}
(E-H_{c})\psi_{c}(\vec{r}) &=& V_{T}(\vec{r})\psi_{f}(\vec{r}),
\\\nonumber
(E-H_{f})\psi_{f}(\vec{r}) &=& V_{T}(\vec{r})\psi_{c}(\vec{r}).
\end{eqnarray}
Then, the confinement wave function $\psi_{c}(\vec{r})$ must be eliminated
from the equations, as it never develops into an asymptotic state.
Thus we get
\begin{equation}
\psi_{f}(\vec{r})=(E-H_{f})^{-1}V_{T}(E-H_{c})^{-1}V_{T}\psi_{f}(\vec{r}).
\end{equation}
From first principles of standard scattering theory, we can conclude that the
factor
\begin{equation}
V_{T}(E-H_{c})^{-1}V_{T}
\end{equation}
acts like an ``effective'' meson-meson potential, which,
if denoted by $V_{MM}$, implies
\begin{equation}
\langle\vec{P}_{f}\mid V_{MM}\mid\vec{P}^{\prime}_{f}\rangle
=\langle\vec{P}_{f}\mid V_{T}(E(\vec{P}_{f})-H_{c})^{-1}V_{T}
\mid\vec{P}^{\prime}_{f}\rangle,
\label{vmm}
\end{equation}
where the total center-of-mass energy (CM) $E$ is given by
\begin{equation}
E(p_{f})=\frac{(\vec{P}_{f})^{2}}{2\mu_{f}}+M_{1}+M_{2},
\end{equation}
with $\mu_{f}$ the reduced mass of the two mesons,
and $P_{f}$ ($P^{\prime}_{f}$) denoting the CM momentum
of the two-meson initial (final) state.

Furthermore, we denote the energy eigenvalue of $H_{c}$ by $E_{nl}$,
i.e.,
\begin{equation}
E_{nl}=\omega (n_{c}+l_{c}+3/2)+M_{q}+M_{\bar{q}},
\label{enl}
\end{equation}
and the corresponding eigensolutions by
$\langle \vec{r}_{c}\mid n_{c},l_{c},m_{c}\rangle$.
By introducing  in Eq.~(\ref{vmm}) a complete set corresponding to this state,
we get
\begin{eqnarray}\nonumber
&& \langle\vec{P}_{f}\mid V_{MM}\mid\vec{P}^{\prime}_{f}\rangle
\\\nonumber
&& =\sum\limits_{n_{c},l_{c},m_{c}}\langle\vec{P}_{f}\mid V_{T}
\mid n_{c},l_{c},m_{c}\rangle\langle n_{c},l_{c},m_{c}
\mid (E(\vec{P}_{f})-H_{c})^{-1}V_{T}\mid\vec{P}^{\prime}_{f}\rangle\\
&& =\sum\limits_{n_{c},l_{c},m_{c}}\langle\vec{P}_{f}
\mid V_{T}\frac{\mid n_{c},l_{c},m_{c}\rangle
\langle n_{c},l_{c},m_{c}\mid}{(E(\vec{P}_{f})-E_{nl})}V_{T}
\mid\vec{P}^{\prime}_{f}\rangle,
\end{eqnarray}
which, upon further introduction of several complete sets
corresponding to the meson-meson configuration space, gives
\begin{eqnarray}\nonumber
&& \langle\vec{P}_{f}\mid V_{MM}
\mid\vec{P}^{\prime}_{f}\rangle\;
=\sum\limits_{n_{c},l_{c},m_{c}}
\int d^{3}r_{f}\int d^{3}r_{f}^{\prime}
\int d^{3}r_{f}^{\prime\prime}\int d^{3}r_{f}^{\prime\prime\prime}
\frac{\langle\vec{P}_{f}\mid\vec{r}_{f}\rangle} 
{(E(\vec{P}_{f})-E_{nl})} \times
\\
&& \times \langle\vec{r}_{f}\mid V_{T}\mid\vec{r_{f}^{\prime\prime}}\rangle
\langle\vec{r_{f}^{\prime\prime}}\mid n_{c},l_{c},m_{c}\rangle
\langle n_{c},l_{c},m_{c}\mid\vec{r_{f}^{\prime\prime\prime}}
\rangle\langle\vec{r_{f}^{\prime\prime\prime}}
\mid V_{T}\mid\vec{r}_{f}^{\prime}\rangle\langle
\vec{r}_{f}^{\prime}\mid\vec{P}^{\prime}_{f}\rangle.
\label{expansion}
\end{eqnarray}

For the transition potential, we take a local delta-shell function of the form
\begin{equation}
\langle\vec{r}_{f}\mid V_{T}\mid\vec{r}_{f}^{\prime}\rangle
=
\frac{\lambda}{\mu_{c}a}\delta (r_{f}- a)
\delta^{3}(\vec{r}_{f}-\vec{r}_{f}^{\prime}).
\label{vt}
\end{equation}

This form of potential has been proven useful
in describing the breaking of the color string
\cite{IJTPGTNO11p179}.
The $\lambda$ and $a$ in Eq.~(\ref{vt}) are the two parameters of the model,
with the former being the coupling of the meson channel
to the quark channel, and the latter an average distance
between the quarks.
The coupling $\lambda$ is varied between 0 and 1 in the present study,
with $\lambda=0$ corresponding to decoupled meson
and quark systems. Since the meson-meson state is considered a plane wave,
decoupling would result in a pure (``bare'') $q\bar{q}$ spectrum.
On the other hand, $\lambda\,\geq$ 1 represents
strong coupling to the meson-meson channel.
The parameter $a$ is taken in the range 3--5~fm.

\begin{figure}[htbp]
\centering
\includegraphics[scale=0.6]{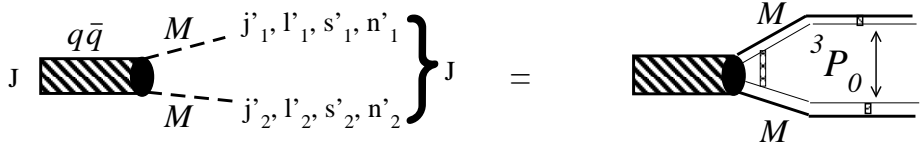}
\caption{The transition vertex from the quark channel
to the two-meson channel.}
\label{fig2}
\end{figure}

Using the above form for $V_{T}$, and the normalization
$\langle\vec{r}\mid\vec{p}\rangle\; = e^{i\vec{p}\cdotp\vec{r}}/(2\pi)^{3/2}$,
Eq.~(\ref{expansion}) becomes
\begin{eqnarray}\nonumber
&& \langle\vec{P}_{f}\mid V_{MM}\mid\vec{P}^{\prime}_{f}\rangle
=\sum\limits_{n_{c},l_{c},m_{c}}
\int\frac{d^{3}r_{f}}{\sqrt{(2\pi)^{3}}}
\int\frac{d^{3}r_{f}^{\prime}}{\sqrt{(2\pi)^{3}}}
e^{-i\vec{P}_{f}\cdotp\vec{r}_{f}}
\frac{\lambda}{\mu_{c}a}\delta(r_{f}-a) \times \\
&& \times \frac{\langle\vec{r}_{f}\mid n_{c},l_{c},m_{c}\rangle
\langle n_{c},l_{c},m_{c}\mid\vec{r}_{f}^{\prime}\rangle}{E(\vec{P_{f}})-E_{nl}}
\frac{\lambda}{\mu_{c}a}\delta(r_{f}^{\prime}-a)
e^{i\vec{P}_{f}^{\prime}\cdotp\vec{r}\,^{\prime}_f}.
\label{vmm2}
\end{eqnarray}
The functions $\langle\vec{r_{f}}\mid n_{c},l_{c},m_{c}\rangle
=Y^{l_f}_{m_f}(\hat {r})\langle|\vec{r_{f}}|\mid n_{c},l_{c},m_{c}\rangle$
represent an overlap wave function of the meson-meson
$\rightarrow\,q\bar{q}$ vertex.
In order to determine this overlap function,
we assume a mechanism for the transition from the meson channel
to the quark channel, or vice versa.
To describe this mechanism, let us consider a confined quark-antiquark
pair with total spin, angular momentum, intrinsic spin, and radial
quantum number $j$, $l_{1}$, $s_{1}$, and $n_{1}$, respectively.
As one usually expects from QCD at low $Q^{2}$,
it is assumed that the string between the quark and the antiquark
breaks with the creation of a new $q\bar{q}$ pair.
This new pair is assumed to have vaccuum quantum number,
corresponding to a $^{2s_{2}+1}l_{2\,j_{2}}=\,^{3\!}P_0$ state.
This yields two $q\bar{q}$ pairs, which then rearrange
to produce two mesons with quantum numbers
$j^{\prime}_{1},l^{\prime}_{1},S^{\prime}_{1},n^{\prime}_{1}$
and $j^{\prime}_{2},l^{\prime}_{2},S^{\prime}_{2},n^{\prime}_{2}$,
respectively (see Fig.~\ref{fig2}).

\begin{figure}[htbp]
\centering
\includegraphics[scale=0.6]{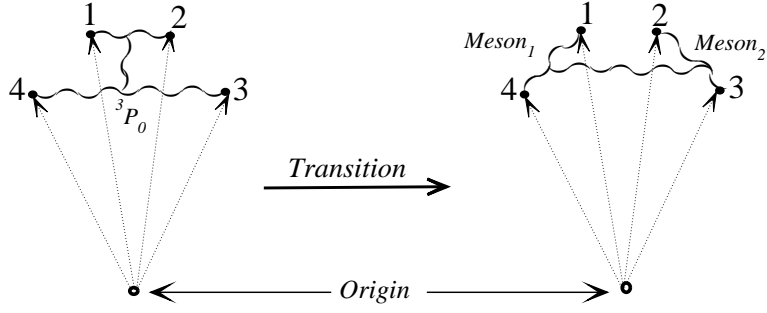}
\vspace{1cm}
\caption{The four quarks, labeled by numbers, before and after the transition.
It is assumed that before the transition the string between
a $q\bar{q}$ pair ($1$ and $2$) breaks to form a new pair ($3$ and $4$),
with  $^{3\!}P_0$ quantum numbers, and the whole system rearranges
so as to produce two mesons made of the quark pairs $1$-$4$ and $3$-$2$.}
\label{fig3}
\end{figure}

Mathematically, the rearrangement of the two $q\bar{q}$ pairs,
one of which is a $^{3\!}P_0$ state and the other having the quantum numbers
of the decaying meson, can be expressed by treating them
as four independent nonrelativistic harmonic oscillators.
Let us label them with numbers,
and assume that system $(1+2)$ represents
the quark-antiquark pair under consideration, while
system $(3+4)$ stands for the
newly created $^{3\!}P_0$ $q\bar{q}$ pair.
This can be treated as a four-body problem,
which can be reduced to a three-body problem
in the global CM system,
by considering the coordinates (momenta) of the CM
of the $(1+2)$, $\vec{r}_{12}(\vec{p}_{12})$, and
$(3+4)$, $\vec{r}_{34}(\vec{p}_{34})$
systems, along with their relative motion
$\vec{r}_{1234}(\vec{p}_{1234})$.
The situation after the transition is described
by assuming that the $(1+4)$ and $(3+2)$ systems
represent the two meson state.
This can be represented diagrammatically,
as shown in Fig.~\ref{fig3}.

The Hamiltonians for the systems, before and after the transition,
can be written in the global CM frame as
\begin{eqnarray}
H_\xscrpt{before}&=&\frac{1}{2}\omega\Bigl
\{ r_{12}^{2}+r_{34}^{2}+r_{1234}^{2}
+p_{12}^{2}+p_{34}^{2}+p_{1234}^{2} \Bigr\},\\
H_\xscrpt{after }  &=&\frac{1}{2}\omega\Bigl
\{ r_{14}^{2}+r_{32}^{2}+r_{1432}^{2}
+p_{14}^{2}+p_{32}^{2}+p_{1432}^{2}\Bigr\},
\end{eqnarray}
where $r_{ij}$ ($p_{ij}$) is the coordinate (momentum) of the $ij$ system,
and $r_{ijkl}$ ($p_{ijkl}$) is the relative coordinate (momentum)
of the CM of the $ij$ and $kl$ systems.
The transition of e.g.\ the coordinates of the four quarks
can be expressed in terms of an orthogonal transformation
matrix $\alpha$
(as explained in Ref.~\cite{ZPC21p291}), i.e.,
\begin{equation}
\left(
\begin{array}{c}
r_{14}\\
r_{32}\\
r_{1432}
\end{array}
\right)
=
\left(
\begin{array}{ccc}
\alpha_{11} & \alpha_{12} & \alpha_{13}\\
\alpha_{21} & \alpha_{22} & \alpha_{23}\\
\alpha_{31} & \alpha_{32} & \alpha_{33}\\
\end{array}
\right)
\left(
\begin{array}{c}
r_{12}\\
r_{34}\\
r_{1234}
\end{array}
\right).
\end{equation}

Thus, if the wave functions of the systems before and after the transition
are written as $\psi^E_{\{n,l,m\}}(r_{12},r_{34},r_{1234})$
and
$\chi^E_{\{n^{\prime},l^{\prime},m^{\prime}\}}(r_{14},r_{32},r_{1432})$,
respectively, then
\begin{eqnarray}\nonumber
&& \langle\psi^E_{\{n,l,m\}}(r_{12},r_{34},r_{1234})\!\mid \; =
\sum\limits_{n^{\prime}, l^{\prime}, m^{\prime}}
\int dr_{14}\int dr_{23}\int dr_{1234}
\\\nonumber
&& \langle\psi^E_{\{n,l,m\}}(r_{12},r_{34},r_{1234})\!
\mid\chi^E_{\{n^{\prime},l^{\prime},m^{\prime}\}}(r_{14},r_{32},r_{1432})
\rangle\langle\chi^E_{\{n^{\prime},l^{\prime},m^{\prime}\}}
(r_{14},r_{32},r_{1432})\mid.
\end{eqnarray}
Now we define a transformation matrix
$D^E_{\{\{n,l,m\},\{n^{\prime},l^{\prime},m^{\prime}\}\}}$ as
\begin{eqnarray}\nonumber
&&D^E_{\{\{n,l,m\},\{n^{\prime},l^{\prime},m^{\prime}\}\}}=
\\\nonumber
&&\int dr_{14}\int dr_{23}\int dr_{1234}
\langle\psi^E_{\{n,l,m\}}(r_{12}, r_{34}, r_{1234})
\mid\chi^E_{\{n^{\prime},l^{\prime},m^{\prime}\}}(r_{14},r_{32},r_{1432})\rangle,
\end{eqnarray}
for which the following analytical expression was obtained
in Ref.~\cite{ZPC21p291}:
\begin{eqnarray}
&&{\cal D}^{E}_{\left\{
\left\{ n,\ell ,m\right\}\, ,\,
\left\{ n^{\prime},{\ell}^{\prime},m^{\prime}\right\}\right\}}
\left(\left\{\bf{r}\right\}\, ;\,\left\{\bf{r}^{\prime}\right\}\right)
\; =\;\\\nonumber
&&
\left (\frac{\pi}{4}\right)^{\frac{1}{2}N(N-1)}
\left\{\prod_{i=1}^{N}(-1)^{n_{i}}
\left[\frac{\Gamma\left(n_{i}+1\right)
\Gamma\left(n_{i}+\ell_{i}+\frac{3}{2}\right)}{2\ell_{i}+1}\right]^{1/2}
\right\}\\\nonumber
&&
\left\{
\prod_{j=1}^{N}(-1)^{n_{j}^{\prime}}
\left[\frac{\Gamma\left(n_{j}^{\prime}+1\right)
\Gamma\left(n_{j}^{\prime}+\ell_{j}^{\prime}+\frac{3}{2}\right)}{2
\ell_{j}^{\prime}+1}\right]^{1/2}
\right\}\sum_{n_{ij},\ell_{ij},m_{ij}}
\nonumber\\ & &
\left[\;
\left\{
\prod_{i=1}^{N}\,
\delta\left(\sum_{j=1}^{N}\left( 2n_{ij}+\ell_{ij}\right),\,
2n_{i}^{\prime}+\ell_{i}^{\prime}\right)\,
\left(\begin{array}{ccc}
\ell_{i1}&\cdots &\ell_{iN}\\
m_{i1}&\cdots & m_{iN}\end{array}\right|\left.
\begin{array}{c}\ell_{i}^{\prime}\\ [10pt] m_{i}^{\prime}\end{array}\right)
\right\}
\right.
\nonumber\\ & &\,\,\,\,\,\left\{
\prod_{j=1}^{N}\,
\delta\left(\sum_{i=1}^{N}\left( 2n_{ij}+\ell_{ij}\right),\,
2n_{j}+\ell_{j}\right)\,
\left(\begin{array}{ccc}
\ell_{1j}&\cdots &\ell_{Nj}\\
m_{1j}&\cdots & m_{Nj}\end{array}\right|\left.
\begin{array}{c}\ell_{j}\\ [10pt] m_{j}\end{array}\right)
\right\}
\nonumber\\ & &\left.\,\,\,\,\,
\left\{
\prod_{i=1}^{N}\,\prod_{j=1}^{N}\,
\left(\alpha_{ij}\right)^{2n_{ij}+\ell_{ij}}\,
\frac{2\ell_{ij}+1}{\Gamma\left(n_{ij}+1\right)
\Gamma\left(n_{ij}+\ell_{ij}+\frac{3}{2}\right)}
\right\}
\;\right] .
\end{eqnarray}
We denote the elements of this transition $D$-matrix by $g_{n}$,
which give the probability of the transition from
a particular meson channel to a particular quark channel, and vice versa.
These $g_{n}$ precisely correspond to the overlap wave functions
$\sqrt{a}/\mu_c\langle|\vec{r_{f}}|\mid n_{c},l_{c},m_{c}\rangle$
in Eq.~(\ref{vmm2}).
Replacing these overlap functions by $\mu_{c}g_{n}/( \sqrt{a})$ turns
Eq.~(\ref{vmm2}) into
\begin{eqnarray}\nonumber
\langle\vec{P}_{f}\mid V_{MM}\mid\vec{P}^{\prime}_{f}\rangle
&=&   \frac{a^4}{(2\pi)^3}\sum\limits_{n_{c},l_{c},m_{c}}
\int  d\Omega_{r_f}
\int  d\Omega_{r{f}^{\prime}}
e^{-i\vec{P}_f \cdotp  a \hat{r}_{f}}
\left( \frac{\lambda}{\mu_{c}a}\right)^2 \frac{\mu_{c}^2g_{n}^2}{ a} \times \nonumber\\
&&\times \frac{Y^{l_f}_{m_f} (\hat{r_f})
Y^{l_f}_{m_f} (\hat{r_f}^\prime)}{E(\vec{P_{f}})-E_{nl}}
e^{i\vec{P}_f^{\prime}\cdotp a \hat{r}_{f}^{\prime}}\nonumber\\
&=&  \frac{\lambda^2 a}{(2\pi)^3}\sum\limits_{n_{c},l_{c},m_{c}}
\int  d\Omega_{r_f}
\int  d\Omega_{r_{f}^{\prime}}
e^{-i\vec{P}_{f}\cdotp  a \hat{r}_{f}}
Y^{l_f}_{m_f} (\hat{r_f})
Y^{l_f}_{m_f} (\hat{r_f}^\prime)  \times \nonumber\\
&&\times \frac{g_{n}^2}{E(\vec{P_{f}})-E_{nl}}
e^{i\vec{P}_{f}^{\prime}\cdotp a \hat{r}_{f}^{\prime}}.\nonumber \\ \label{simplify}
\end{eqnarray}
Using the standard relations for spherical harmonics
\begin{equation}
\int d\Omega_{r_f}  e^{-i\vec{P}_f.\hat{r}_f a} Y_{m_f}^{l_f } (\hat{r}_f)
=  (-i)^l_f 4 \pi j_l (P_f a) Y_{m_f}^{l_f } (\hat{P}_f)
\end{equation}
in Eq.~({\ref{simplify}}), we get \\[11pt]
$\langle\vec{P}_{f}\mid V_{MM}\mid\vec{P}^{\prime}_{f}\rangle \; =$ \\[-13pt]
\begin{eqnarray}\nonumber
&=&  \frac{\lambda^2 a}{(2\pi)^3}\sum\limits_{n_{c},l_{c},m_{c}}
(-i)^{l_f} 4 \pi j_l (P_f a) Y_{m_f}^{l_f } (\hat{P}_f)
(i)^{l_f} 4 \pi j_l (P_f^\prime a) Y_{m_f}^{* l_f } (\hat{P}^\prime_f)
\frac{g_{n}^2}{E(\vec{P_{f}})-E_{nl}}\nonumber\\
&=&  \frac{\lambda^2 a}{(2\pi)^3}\sum\limits_{n_{c},l_{c},m_{c}}
(-i^2)^{l_f} (4 \pi)^2 j_l (P_f a)  j_l (P_f^\prime a) Y_{m_f}^{l_f } (\hat{P}_f)
Y_{m_f}^{* l_f } (\hat{p}^\prime_f)
\frac{g_{n}^2}{E(\vec{P_{f}})-E_{nl}}\nonumber\\
&=&  \frac{\lambda^2 a}{(2\pi)^3}\sum\limits_{n_{c},l_{c}}
(-i^2)^{l_f} (4 \pi)^2 j_l (P_f a)  j_l (P_f^\prime a)
\frac{2l_f+1}{4\pi} \mathbb{P}_{l_f}(\hat{P_{f}}\cdot\hat{P}_{f}^{\prime})
\frac{g_{n}^2}{E(\vec{P_{f}})-E_{nl}}
\nonumber\\
&=& \frac{ \lambda^2 a} {2\pi^2}
\sum\limits_{l_c=0}^{\infty}(2l_{f}+1)
\mathbb{P}_{l_{f}}(\hat{P_{f}}\cdot\hat{P}_{f}^{\prime} )
j_l(P_{f}a) j_l(P_{f}^{\prime} a)\sum\limits_{n_c=0}^{\infty}
\frac{g_{n}^{2}}{E(\vec{P_{f}})-E_{nl}},
\end{eqnarray}
where $\mathbb{P}_l (x )$ is the Legendre polynomial.

Using this potential, a simple closed-form expression for the $S$-matrix
can be obtained, if only one confined and one free channel are considered
(see Appendices A.1--A.5
of Ref.~\cite{IJTPGTNO11p179} for  a detailed derivation), viz.\
\begin{equation}
S_{l_{f}}(E) = 1 - 2i\frac{2a\lambda^{2}\sum\limits_{n=0}^\infty
\dfrac{g_{n}^{2}}{E(\vec{P_{f}})-E_{nl}}\mu_{f}P_{f}j_{l_{f}}
(P_{f}a) h_{l_{f}}^{1}(P_{f}a)}
{1+2ia\lambda^{2}\sum\limits_{n=0}^\infty
\dfrac{g_{n}^{2}}{E(\vec{P_{f}})-E_{nl}}
\mu_{f}P_{f}j_{l_{f}}(P_{f}a) h_{l_{f}}^{1}(P_{f}a)}.
\label{smat}
\end{equation}
An exact solution for the $S$-matrix can be derived in the most general 
multichannel case as well, resulting in a matrix expression with a 
similar structure \cite{PRD80p094011}.

\begin{figure}[htbp]
\centering
\includegraphics[scale=0.45,angle=-90]{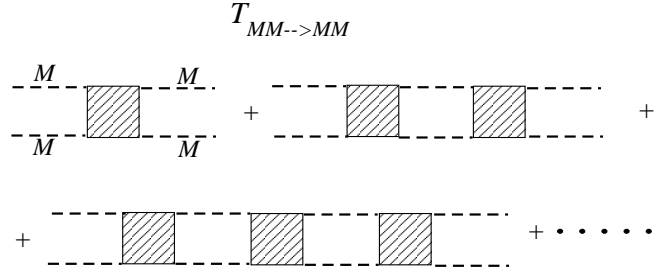}
\caption{The full scattering amplitude for two mesons,
with the shaded boxes given by Fig.~\ref{fig1}.}
\label{fig4}
\end{figure}

The full scattering amplitude can be depicted diagrammatically
as shown in Fig.~\ref{fig4}, where the shaded boxes represent
the effective meson-meson potential depicted in Fig.~\ref{fig1}.

In order to find resonances in non-exotic meson-meson systems,
we search for zeros in the denominator of the $S$-matrix (Eq.~(\ref{smat})),
which correspond to poles in the complex-energy plane. Then we vary the
parameters $\lambda$ and $a$ to fit the experimental data for a particular
case. We also study the pole movements in the complex plane, as we change
the parameter $\lambda$, in order to get more
physical insight into the model, and to evaluate the role of meson loops
coupled to quark channels in the generation of resonances.

\section{Perturbative formalism}
\label{Perturbative}

The main purpose of this paper is to show that predictions
of resonances poles based on perturbative calculations in standard quark
models can be very misleading. In order to demonstrate this in a quantitative
way, we now construct a perturbative scheme for the formalism described in the
previous section.

As explained above, the $\lambda$ in our formalism
is the coupling of the meson-meson $\leftrightarrow\, q\bar{q}$ vertex.
The term corresponding to $\lambda^{2}$ thus represents
the lowest-order meson-meson interaction
(meson-meson $\rightarrow\, q\bar{q} \, \rightarrow\,$ meson-meson ).
The position of a pole in a particular case of meson-meson scattering,
above threshold, can be expanded perturbatively in terms of $\lambda^{2}$ as
\begin{equation}
E_{m}^\xscrpt{pole}= E_{m} +\lambda^{2} E_{m}^\xscrpt{LO}
+\lambda^{4} E_{m}^\xscrpt{NLO}
+\lambda^{6} E_{m}^\xscrpt{NNLO}+\ldots .
\label{empole}
\end{equation}
The first term of this series corresponds to the confinement pole,
which is what we should get from the model if the coupling of the quark pair
to the meson channels vanished.
The second term is the leading-order term in $\lambda^{2}$.
The denominator of the $S$-matrix (Eq.~(\ref{smat})), written
up to leading order in $\lambda^{2}$
around the $m$th confinement pole, becomes
\begin{equation}
1+2ia\lambda^{2}\frac{g_{m}^{2}}
{E_{m}+\lambda^{2} E_{m}^\xscrpt{LO}-E_{m}}
\mu k j_l(ka) h_l^{1}(ka) = 0,
\end{equation}
which then yields
\begin{equation}
E_{m}^\xscrpt{LO}= - 2ia g_{m}^{2}\mu k j_l(ka) h_l^{1}(ka).
\end{equation}
So the pole position (Eq.~(\ref{empole})),
in lowest-order approximation, is
\begin{equation}
E_{m}^\xscrpt{pole}\approx E_{m}-2ia g_{m}^{2}\mu k j_l(ka) h_l^{1}(ka).
\end{equation}

We now expand the denominator of Eq.~(\ref{smat})
to higher order in $\lambda^{2}$.
In order to do so, we define
\begin{equation}
f(\lambda^{2}) = 2ia \mu k j_l(ka) h_l^{1}(ka),\label{ffn}
\end{equation}
such that Eq.~(\ref{smat}) becomes
\begin{equation}
S_l (E) = 1 -\frac{2\lambda^{2}f(\lambda^{2})
\sum\limits_{n=0}^\infty\dfrac{g_{n}^{2}}{E-E_{n}}}
{1+\lambda^{2}f(\lambda^{2})
\sum\limits_{n=0}^\infty\dfrac{g_{n}^{2}}{E-E_{n}}}.
\label{smat2}
\end{equation}
Now we expand the function $f$ around the energy $E=E_{m}$, i.e.,
around $\lambda = 0$, as
\begin{eqnarray}\nonumber
f(E_{m}) &=& f(E_{m}) +\lambda^{2}
\frac{\partial f}{\partial\lambda^{2}}
\Big\vert_{\lambda=0}+
\frac{1}{2}\lambda^{4}\frac{\partial^{2} f}
{\partial(\lambda^{2})^{2}}\Big\vert_{\lambda=0}
+\frac{1}{6}\lambda^{6}
\frac{\partial^{3} f}{\partial(\lambda^{2})^{3}}\Big\vert_{\lambda=0}
\nonumber\\ &+&\ldots\nonumber\\
&=& f(E_{m})
+\lambda^{2}\frac{\partial f}{\partial E}\frac{\partial E}
{\partial\lambda^{2}}\Big\vert_{\lambda=0}
+\frac{1}{2}\lambda^{4}
\left[
\frac{\partial^{2} f}{\partial E^{2}}\left(\frac{\partial E}
{\partial\lambda^{2}}\right)^{2}\right.
\nonumber\\
&+&\left.
\frac{\partial f}{\partial E}\frac{\partial^{2} E}{\partial\lambda^{2}}
\right]_{\lambda=0}+\frac{1}{6}\lambda^{6}\left[
\frac{\partial^{3} f}{\partial E^{3}}\left(\frac{\partial E}
{\partial\lambda^{2}}\right)^{3}
+3\frac{\partial^{2} f}{\partial E^{2}}\frac{\partial E}
{\partial\lambda^{2}}
\frac{\partial^{2} E}{\partial(\lambda^{2})^{2}}\right.
\\
&+&\left.
\frac{\partial f}{\partial E}
\frac{\partial^{3} E}{\partial(\lambda^{2})^{3}}
\right]_{\lambda=0}+
\ldots .
\label{fexp}
\end{eqnarray}
From Eq.~(\ref{empole}), we have
$E_{m}^\xscrpt{pole}|_{\lambda=0}=E_m$, and
\begin{equation}\nonumber
\frac{\partial E_{m}^\xscrpt{pole}}{\partial\lambda^{2}}
\Big\vert_{\lambda=0}= E_{m}^\xscrpt{LO},\,\,
\frac{\partial^{2} E_{m}^\xscrpt{pole}}{\partial(\lambda^{2})^{2}}
\Big\vert_{\lambda=0}= 2E_{m}^\xscrpt{NLO},\,\,
\frac{\partial^{3} E_{m}^\xscrpt{pole}}{\partial(\lambda^{2})^{3}}
\Big\vert_{\lambda=0}= 6E_{m}^\xscrpt{NNLO}, \,\,\ldots .
\end{equation}
Using these relations in Eq.~(\ref{fexp}), we get
\begin{eqnarray}\label{fexp2}
f(E_{m}) &=&
f(E_{m})+\lambda^{2} E_{m}^\xscrpt{LO}
\frac{\partial f}{\partial E}\Big\vert_{E=E_{m}}
\nonumber\\
&+&\frac{1}{2}\lambda^{4}
\left[\left( E_{m}^\xscrpt{LO}\right)^{2}
\frac{\partial^{2} f}{\partial E^{2}}\Big\vert_{E=E_{m}}
+ 2 E_{m}^\xscrpt{NLO}\frac{\partial f}{\partial E}\Big\vert_{E=E_{m}}
\right]\nonumber\\
&+&\frac{1}{6}\lambda^{6}\left[\left( E_{m}^\xscrpt{LO}\right)^{2}
\frac{\partial^{3} f}{\partial E^{3}}\Big\vert_{E=E_{m}}
+6 E_{m}^\xscrpt{NLO}E_{m}^\xscrpt{LO}
\frac{\partial^{2} f}{\partial E^{2}}\Big\vert_{E=E_{m}}
\right]
\nonumber\\
&+& 6 E_{m}^\xscrpt{NNLO}\frac{\partial f}{\partial E}\Big\vert_{E=E_{m}}+
\ldots .
\end{eqnarray}
It remains to consider the remaining part of the denominator, namely
\begin{equation}
\lambda^{2}\sum\limits_{n=0}^\infty\frac{g_{n}^{2}}{E - E_{n}}=
\lambda^{2}\sum\limits_{n\neq m}^\infty\frac{g_{n}^{2}}{E - E_{n}}+
\lambda^{2}\frac{g_{m}^{2}}{E- E_{m}},
\end{equation}
which we expand in a series in $\lambda^{2}$,
around $\lambda=0$ (so at $E=E_{m}$),
in a similar way as the function $f$, to obtain
\begin{eqnarray}\nonumber
&&\lambda^{2}\sum\limits_{n=0}^\infty\frac{g_{n}^{2}}{E - E_{n}}=
\lambda^{2}\sum\limits_{n\ne m}\frac{g_{n}^{2}}{E_{m}- E_{n}}-
\lambda^{4} E_{m}^\xscrpt{LO}\sum\limits_{n\ne m}
\frac{g_{n}^{2}}{(E_{m}- E_{n})^{2}}
\\\nonumber
&&+\lambda^{6}\left\{ (E_{m}^\xscrpt{LO})^{2}\sum\limits_{n\ne m}
\frac{g_{n}^{2}}{(E_{m}- E_{n})^{3}}
-E_{m}^\xscrpt{NLO}\sum\limits_{n\ne m}
\frac{g_{n}^{2}}{(E_{m}- E_{n})^{2}}\right\}
\\\nonumber
&&+g_{m}^{2}\left[\frac{1}{E_{m}^\xscrpt{LO}}-\lambda^{2}
\frac{E_{m}^\xscrpt{NLO}}{(E_{m}^\xscrpt{LO})^{2}}
+\lambda^{4}\left\{\frac{(E_{m}^\xscrpt{NLO})^{2}}
{(E_{m}^\xscrpt{LO})^{3}}-
\frac{E_{m}^\xscrpt{NNLO}}{(E_{m}^\xscrpt{LO})^{2}}\right\}
\right]\\ \label{expn3}&&
+g_{m}^{2}\left[
-\lambda^{6}\left\{\frac{(E_{m}^\xscrpt{NLO})^{3}}
{(E_{m}^\xscrpt{LO})^{4}}-
2\frac{E_{m}^\xscrpt{NLO}*E_{m}^\xscrpt{NNLO}}
{(E_{m}^\xscrpt{LO})^{3}}
+\frac{E_{m}^\xscrpt{N$^{3}$LO}}{(E_{m}^\xscrpt{LO})^{2}}
\right\}
\right] +\ldots.
\end{eqnarray}

Upon multiplying Eqs.~(\ref{fexp2}) and (\ref{expn3}),
we get an expansion in $\lambda^{2}$ for the pole position, viz.\

\begin{eqnarray}\nonumber
&0& = 1+{g_{m}^{2}}{E_{m}^\xscrpt{LO}}f(E_{m})\\\nonumber
&+&\lambda^{2}\left\{ \left(\sum\limits_{n\ne m}
\frac{g_{n}^{2}\left( E_{m}^\xscrpt{LO}\right)^{2} }{E_{m}- E_{n}}
- g_{m}^{2} E_{m}^\xscrpt{NLO}\right)
\frac{ f(E_{m})}{\left( E_{m}^\xscrpt{LO}\right)^{2} }
+ g_{m}^{2}\frac{\partial f}{\partial E}\Big\vert_{E=E_{m}}\right\}
\\\nonumber
&+&\lambda^{4}\left\{ \left(-\sum\limits_{n\ne m}\frac{g_{n}^{2}
\left( E_{m}^\xscrpt{LO}\right)^{4} }{(E_{m}- E_{n})^{2}}+ g_{m}^{2}
\left\{ \left( E_{m}^\xscrpt{NLO}\right)^{2}
-E_{m}^\xscrpt{LO}E_{m}^\xscrpt{NNLO}\right\}
\right)\frac{ f(E_{m})}{\left( E_{m}^\xscrpt{LO}\right)^{3} }\right.
\\\nonumber
&+&\left.
\left(\sum\limits_{n\ne m}\frac{g_{n}^{2}
E_{m}^\xscrpt{LO}}{E_{m}- E_{n}}\right)
\frac{\partial f}{\partial E}\Big\vert_{E=E_{m}}
+\frac{1}{2} g_{m}^{2}
E_{m}^\xscrpt{LO}\frac{\partial^{2} f}{\partial E^{2}}
\Big\vert_{E=E_{m}}\right\}
\\\nonumber
&+&\lambda^{6}\left\{ \left(\sum\limits_{n\ne m}\frac{g_{n}^{2}
\left( E_{m}^\xscrpt{LO}\right)^{6} }{(E_{m}- E_{n})^{3}}
-\sum\limits_{n\ne m}
\frac{g_{n}^{2}\left( E_{m}^\xscrpt{LO}\right)^{4}E_{m}^\xscrpt{NLO}}
{(E_{m}- E_{n})^{2}}\right.\right.
\\\nonumber
&-&\left.\left.g_{m}^{2}\left\{
\left( E_{m}^\xscrpt{NLO}\right)^{3}
-2E_{m}^\xscrpt{NLO}E_{m}^\xscrpt{NNLO}E_{m}^\xscrpt{LO}
+E_{m}^\xscrpt{N$^{3}$LO}
\left( E_{m}^\xscrpt{LO}\right)^{2}\right\}\right)
\frac{ f(E_{m})}{\left( E_{m}^\xscrpt{LO}\right)^{4} }\right.
\\\nonumber
&+&\left( -\sum\limits_{n\ne m}\frac{g_{n}^{2}\left( E_{m}^\xscrpt{LO}
\right)^{2} }{(E_{m}- E_{n})^{2}}+ \sum\limits_{n\ne m}
\frac{g_{n}^{2} E_{m}^\xscrpt{NLO}}{E_{m}- E_{n}}\right)
\frac{\partial f}{\partial E}\Big\vert_{E=E_{m}}+\\\nonumber
&+&\left.
\frac{1}{2}\left(\sum\limits_{n\ne m}\frac{g_{n}^{2}
\left( E_{m}^\xscrpt{LO}\right)^{2} }
{E_{m}- E_{n}}+  g_{m}^{2} E_{m}^\xscrpt{NLO}\right)
\frac{\partial^{2} f}{\partial E^{2}}\Big\vert_{E=E_{m}}\right.
\\\nonumber
&+&\left.
\frac{1}{6}g_{m}^{2}\left( E_{m}^\xscrpt{LO}\right)^{2}
\frac{\partial^{3} f}{\partial E^{3}}\Big\vert_{E=E_{m}}\right\}.
\end{eqnarray}
Solving this equation order by order in $\lambda^{2}$,
we obtain the expressions for e.g.\
the pole position to lowest order, next-to-lowest order,
and next-to-next-to-lowest order as
\begin{eqnarray}
&&E_{m}^\xscrpt{LO}= - g_{m}^{2} f(E_{m}),\label{one}\\
&&E_{m}^\xscrpt{NLO}= g_{m}^{4} f(E_{m})\frac{\partial f}{\partial E}
\Big\vert_{E=E_{m}}+ g_{m}^{2} f^{2}(E_{m})\sum\limits_{n\ne m}
\frac{g_{n}^{2} }{E_{m}- E_{n}},\label{two}
\end{eqnarray}
\begin{eqnarray}\nonumber
&&E_{m}^\xscrpt{NNLO}= g_{m}^{4} f^{3}(E_{m})\sum\limits_{n\ne m}
\frac{g_{n}^{2} }{(E_{m}- E_{n})^{2}}- g_{m}^{2} f^{3}(E_{m})\times
\\\nonumber &&\left(\sum\limits_{n\ne m}
\frac{g_{n}^{2} }{E_{m}- E_{n}}\right)^{2} -3 g_{m}^{4} f^{2}(E_{m})
\frac{\partial f}{\partial E}\Big\vert_{E=E_{m}}
\left(\sum\limits_{n\ne m}\frac{g_{n}^{2} }{E_{m}- E_{n}}\right)\\
&&-\frac{1}{2} g_{m}^{6} f^{2}(E_{m})\frac{\partial^{2} f}{\partial E^{2}}
\Big\vert_{E=E_{m}}- g_{m}^{6} f(E_{m})
\left(\frac{\partial f}{\partial E}\Big\vert_{E=E_{m}}\right)^{2}.
\label{three}
\end{eqnarray}
Similarly, one can obtain the expressions for even higher-order contributions.

\section{Results and discussion}
To test the validity of the perturbative calculus, we now choose a few
concrete examples of meson-meson systems, namely $K\pi$ in $P$ wave,
$D\bar{D}$ in $P$ wave, and $DK$ in $S$ wave.
First we compute the scattering poles
in these systems by using the exact formalism,
explained in Sect.~\ref{Formalism}.
In particular, we study the  $S$-matrix (Eq.~(\ref{smat})) pole positions 
in the complex-energy plane, as a function of the coupling $\lambda$.
The resulting pole trajectories behave 
as expected from general considerations:
\begin{itemize}
\item{
The unperturbed, or bare,
quark-antiquark spectrum is given by Eq.~(\ref{enl}).
For small coupling to the scattering sector, we must find
resonance poles close to the levels of this spectrum.}
\item{
For larger coupling, the resonances are expected
to acquire larger widths,
so the pole positions must get larger
(negative) imaginary parts.
However, associated with larger widths are, in general, larger mass shifts.
As a consequence, also the real parts of the pole positions
will deviate substantially from the levels of the
bare $q\bar{q}$ spectrum.}
\item{
Sometimes, a further increased coupling may lead to a sufficiently large
mass shift so as to push the pole below the scattering threshold. For such a
situation, we expect a bound state.
Since a bound state has zero width, the pole should eventually end up on
the real-energy axis.}
\end{itemize}

Next, we study the same cases, but now
using the perturbative formalism derived in
Sect.~\ref{Perturbative}, up to
fourth order in $\lambda^2$.
This way we can compare exact and perturbative results
for the real and imaginary mass shifts due to meson loops.
The latter quantities are related to predictions for
the central resonance masses and resonance widths.
We will find that, for moderate to large couplings, the perturbative results
strongly deviate from and do not converge towards the exact ones.
In particular, the expected behavior of poles
below the scattering threshold, namely to move towards or along the
real-energy axis, cannot be reproduced at all by the studied perturbative
expansions.

In all cases we will employ the quark masses
$m_{n}\equiv m_{u}=m_{d}=0.406$~GeV, $m_{s}=0.508$~GeV,
$m_{c}=1.562$~GeV, as well as the universal oscillator frequency
 $\omega =0.19$~GeV, all determined in the model of Ref.~\cite{PRD27p1527}.
The values of the parameters $a$ and $\lambda$ are given separately for each
case in the following discussion.

The parameter $a$ in Eq.~(\ref{vt})
describes the average distance at which quark-pair
creation or annihilation takes place,
leading to the two-meson decay of a meson.
For the case of $K\pi$, which is a nonstrange-strange
flavor combination, we choose here $a=2.534$ GeV$^{-1}$.
This is smaller than the value $a=3.2$ GeV$^{-1}$,
which would be used in a multichannel fit to several mesons.
The reason for this discrepancy is that one meson-meson channel,
namely $K\pi$, has to mimic the effect of a multichannel treatment.
In order to nonetheless obtain the $K^{\ast}(892)$ pole
at a reasonable position, we have to adjust the parameter $a$.
In the $D\bar{D}$ and $DK$ cases, we use here values
close to the ones used in a multichannel calculation \cite{PRD27p1527},
namely $a=1.72$~GeV$^{-1}$ and $a=2.5$~GeV$^{-1}$, respectively.
The price we pay is that, in the $D\bar{D}$ case, the $\psi(2S)$
pole does not come out at 3.686~GeV, but about 40--50~MeV higher.

For the parameter $\lambda$, which is the universal overall
three-meson coupling constant, we could have used, after scaling,
the same value in all three cases.
However, scaling was not carried out \cite{PRL91p012003} in the $DK$
case, which makes that the pole now greatly overshoots
the $D_{s0}(2317)$ position for $\lambda =1$.
With the correct scaling, it would end up roughly 20 MeV too high.

One might argue that only one value of the coupling $\lambda$
describes the physical situation, so that other values are not relevant.
However, as analyticity has proven in the past
to be a powerful tool for constructing scattering amplitudes,
the trajectories of their poles are also
a strong indication for the correctness of their dependence
on other parameters.
In the following, we will show that perturbative expansions,
even to higher orders, only have a very limited range of validity,
and do not cover the realistic case of large couplings
\cite{ZPC30p615,PRD59p074001,ARXIV10080466}
in strong interactions.

\subsection{$P$-wave $K\pi$ scattering}
\label{Kpiscattering}

\begin{figure}[htbp]
\centering
\includegraphics[scale=0.5]{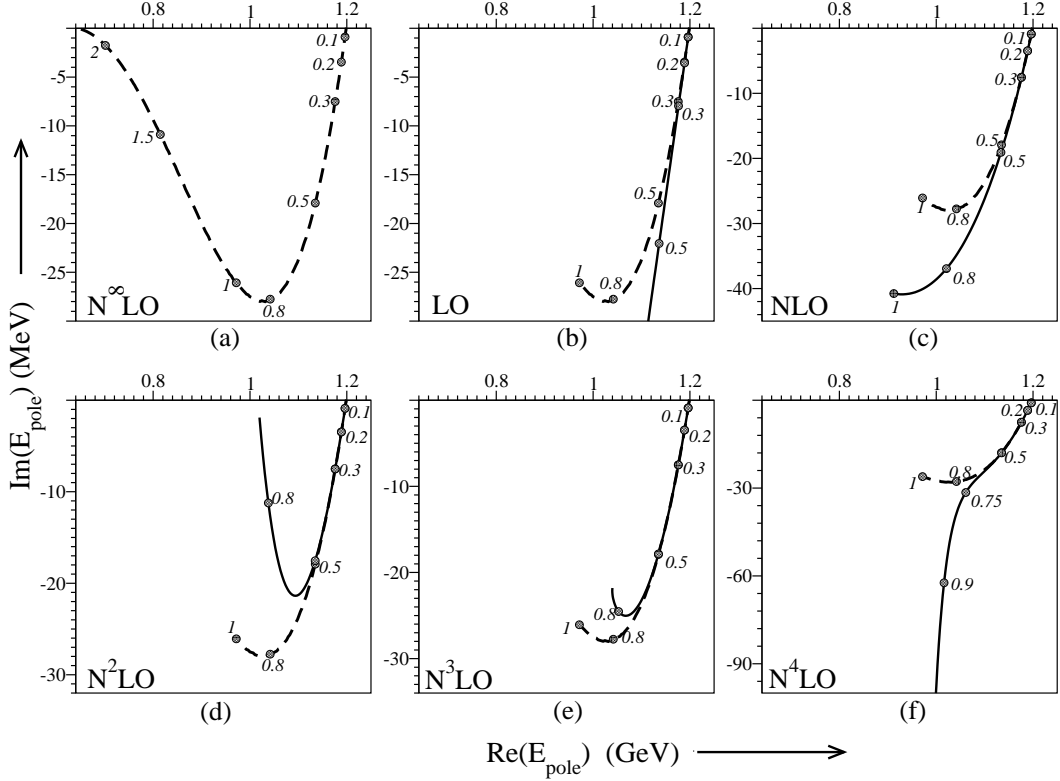}
\caption{\small
The $P$-wave $K\pi$ resonance pole positions
in the complex-energy plane,
corresponding to the second Riemann sheet,
under variation of the coupling $\lambda$.
The dashed curves in all the plots for N$^{n}$LO ($n=0$, $\dots$, 4)
are the same as the ones shown for N$^{\infty}$LO (labeled (a)).
The solid curves are the results obtained
from the  perturbative approximations, viz.\
(b) leading (N$^{0}$LO, Born) term,
(c) next-to-leading (N$^{1}$LO),
(d) next-to-next-to-leading (N$^{2}$LO),
(e) (next-to)$^{3}$-leading (N$^{3}$LO)
and (f) (next-to)$^{4}$-leading (N$^{4}$LO) orders, respectively.}
\label{figKpiP}
\end{figure}

The $K^{\ast}(892)$ resonance is well described by a Breit-Wigner
resonance in $P$-wave $K\pi$ scattering,
with central mass and resonance width of about 892 MeV
and 50 MeV, respectively.
Hence, we expect a pole in the $S$ matrix of Eq.~(\ref{smat})
for an $S$-wave nonstrange-strange quark-antiquark system,
coupled to a $P$-wave kaon-pion meson-meson system.
For the couplings $g_{n}$ we find 
\cite{ZPC21p291} in this case
\begin{equation}
g_{n} = 2^{-n}\left(\frac{2n+3}{3}\right)^{1/2}.
\label{VPP}
\end{equation}
Scattering poles are obtained by studying the zeros
of the denominator in the expression of Eq.~(\ref{smat}).
In Fig.~\ref{figKpiP}(a), we depict
the $S$-matrix pole positions 
for a range of  $\lambda$ values
varying from 0 to just over 2.

In the limit of vanishing coupling,
one expects to find the poles at the bare masses
of the quark-antiquark system, as given by Eq.~(\ref{enl}).
We obtain from Eq.~(\ref{enl}) the value $E_{00}=1.199$ GeV for the
ground-state bare mass, which indeed corresponds to the limit of
vanishing $\lambda$ along the dashed curve in Fig.~\ref{figKpiP}(a).
For larger couplings, we observe that the imaginary part
of the pole position vanishes at the $K\pi$ threshold.
This was to be expected, since a large coupling results in a bound
state below the $K\pi$ threshold, which of course has a zero width.
The shape of the pole trajectory near the $K\pi$ threshold
is in accordance with theory for poles in $P$-wave scattering and also
in higher partial waves \cite{Taylor,LNP211p331}.
For $S$-wave scattering the pole behavior is different,
as we will see in Sect.~\ref{DKscattering},
but again in agreement with theory 
\cite{Taylor,LNP211p331}.

The value $\lambda=1$ corresponds to the physical pole,
as it roughly reproduces the characteristics of the $K^*$ (892)
resonance. In the present simplified model, the pole comes out
at $(0.972-i0.026)$~GeV, as shown in Fig.~\ref{figKpiP}(a).

The coefficients of the perturbative expansion
(Eq.~(\ref{empole})) are collected
in Table~\ref{Kpicoefficients},
for the case of $P$-wave $K\pi$ scattering.
\begin{table}[htbp]
\begin{center}
\begin{tabular}{||l|r||}
\hline\hline & \\ [-5pt]
coefficient & value (GeV)\\ [5pt]
\hline & \\ [-5pt]
$E_{0}$ & (1.199, 0.)\\
$E_{0}^{\xscrpt{N}^{0}\xscrpt{LO}}$ & (-0.249080686,-0.0878366188)\\
$E_{0}^{\xscrpt{N}^{1}\xscrpt{LO}}$ & (-0.0435913117,0.0471697828)\\
$E_{0}^{\xscrpt{N}^{2}\xscrpt{LO}}$ & (0.0631440181,0.0973648258)\\
$E_{0}^{\xscrpt{N}^{3}\xscrpt{LO}}$ & (0.0869057944,-0.0785515047)\\
$E_{0}^{\xscrpt{N}^{4}\xscrpt{LO}}$ & (-0.067527691,-0.0632886834)\\
\hline\hline
\end{tabular}
\end{center}
\caption[]{Coefficients of the perturbative expansion
given in Eq.~(\ref{empole}), concerning the pole positions of the
ground-state pole in $P$-wave $K\pi$ scattering.}
\label{Kpicoefficients}
\end{table}
In Figs.~\ref{figKpiP}(b--f) we depict
the perturbative pole trajectories for the bare nonstrange-strange
$q\bar{q}$ state at 1.199~GeV. Shown are the curves
for the lowest-order (Born) term
($E_{0}^{\xscrpt{N}^{0}\xscrpt{LO}}$)
and for the next few higher-order terms
($E_{0}^{\xscrpt{N}^{1}\xscrpt{LO}}$, \ldots), respectively, up to 
fourth order in $\lambda^2$.
We find that the Born term gives satisfactory 
pole positions for overall couplings up to $\lambda\approx0.3$.
At each higher order, the perturbative pole positions,
i.e., the central masses and the widths
of the $K^{\ast}(892)$ resonance, are better and better determined,
up to $\lambda\approx0.75$ for the fourth-order approximation.
However, thereabove things go terribly wrong, and all approximations
completely fail to reproduce the physical pole at $\lambda=1$.

So we are forced to conclude that perturbation theory is unreliable
to describe the $K^{\ast}(892)$ resonance.
Moreover, we should add that these higher-order perturbative calculations
are much more tedious than just finding the exact solution for the coupled
quark-antiquark and meson-meson system.

\subsection{$P$-wave $D\bar{D}$ scattering}
\label{DDscattering}

Let us next consider the $D\bar{D}$ system,
which has been studied already a long time ago \cite{PRD21p772},
using the model described in Sect.~\ref{Formalism}.
In Ref.~\cite{PRD21p772} it was shown that the $P$-wave $D\bar{D}$ channel,
together with higher open-charm channels, can transform the bare vector
charmonium spectrum into the physical one. In particular, the pole stemming 
from the first radial excitation comes out very close to the
$\psi(2S)$(3686) state, which turns out to contain a significant
$D\bar{D}$ component, besides $c\bar{c}$ of course.

\begin{figure}[htbp]
\centering
\includegraphics[scale=0.5]{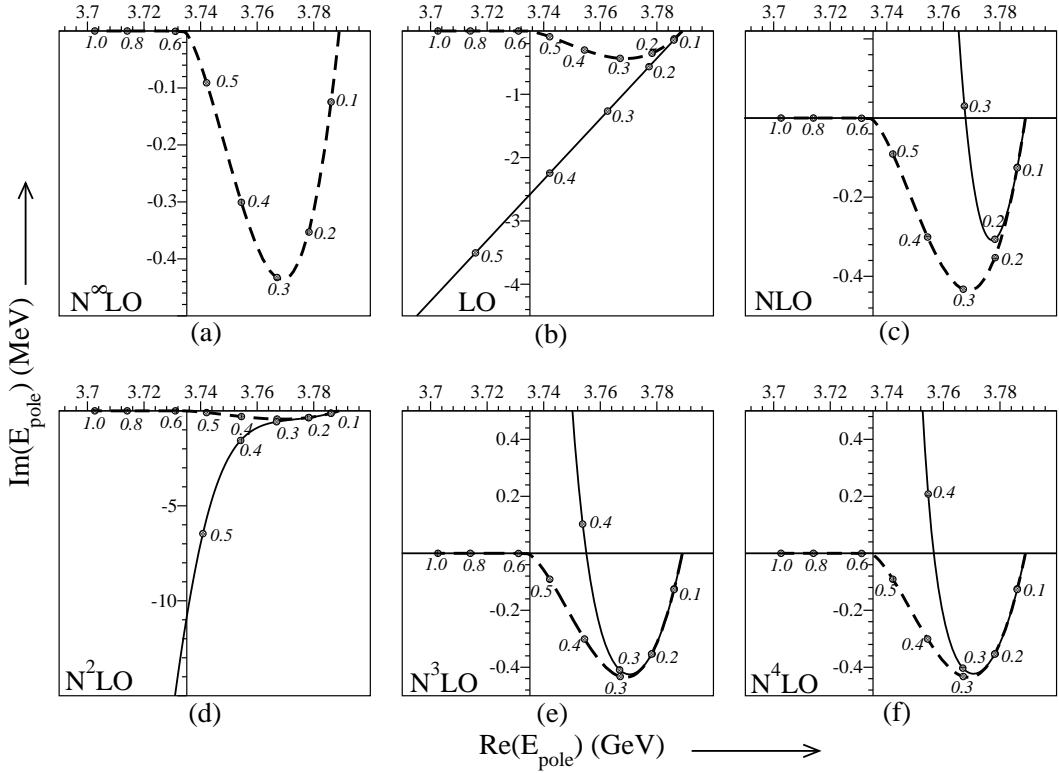}
\caption{\small
$P$-wave $D\bar{D}$ pole positions
in the complex-energy plane
(corresponding to the first and second Riemann sheets),
under variation of the  coupling $\lambda$.
The dashed curves in all the plots for N$^{n}$LO ($n=0$, $\dots$, 4)
are the same as the one for N$^{\infty}$LO (labeled (a)).
The solid  curves represent the perturbative results:
(b) leading (N$^{0}$LO, Born) term,
(c) next-to-leading (N$^{1}$LO),
(d) next-to-next-to-leading (N$^{2}$LO),
(e) (next-to)$^{3}$ (N$^{3}$LO),
and (f) (next-to)$^{4}$-leading (N$^{4}$LO) orders, respectively.}
\label{figDD}
\end{figure}

The couplings $g_{n}$ in this case are
again given by the vector $\leftrightarrow$ pseudoscalar-pseudo\-scalar
vertex, for which we use the same expression as in Eq.~(\ref{VPP}).
The parameter $a$ is now taken at 0.34~fm.
Using these inputs, the $S$-matrix (Eq.~(\ref{smat}))
is calculated, and we search for poles on the second Riemann sheet.
We present the results of our calculation in Fig.~\ref{figDD}(a),
which depicts the complex-energy plane around the mass
of the $\psi(2S)$(3686).
The dashed line in Fig.~\ref{figDD}(a)
corresponds to the movement of the pole in the complex plane
between the $D\bar{D}$ threshold
and the first radial excitation of the $J\!/\!\psi$,
as $\lambda$ is varied between the limiting values 0 and 1.
As expected, when the coupling is very small,
we find a pole close to the first radial excitation
of the confined $c\bar{c}$ spectrum of the model,
i.e., near $E_{1} =\omega (2+3/2 )+2m_{c}=$ 3.789 GeV (Eq.~(\ref{enl})).
As $\lambda$ is increased to 0.1,
the real part of the pole becomes $\Re\mbox{e}(E)\simeq3.77$~GeV
Finally, for $\lambda\simeq1$,
the pole is found below the $D\bar{D}$ threshold, very close to 3.7 GeV,
which should corresponds to the physical $\psi(2S)$(3686).

Next we show the pole positions obtained in
the perturbative expansion, viz.\ from Eqs.~(\ref{one}--\ref{three}))
and similar expressions up to fourth order in $\lambda^2$.
The coefficients of Eq.~(\ref{empole})
are collected in Table~\ref{DDcoefficients}.
\begin{table}[htbp]
\begin{center}
\begin{tabular}{||l|r||}
\hline\hline & \\ [-5pt]
coefficient & value (GeV)\\ [5pt]
\hline & \\ [-5pt]
$E_{0}$ & (3.789, 0.)\\
$E_{0}^{\xscrpt{N}^{0}\xscrpt{LO}}$ & (-0.291265968,-0.0139626856)\\
$E_{0}^{\xscrpt{N}^{1}\xscrpt{LO}}$ & (0.587237403,0.157827196)\\
$E_{0}^{\xscrpt{N}^{2}\xscrpt{LO}}$ & (-0.763203829,-0.81585325)\\
$E_{0}^{\xscrpt{N}^{3}\xscrpt{LO}}$ & (-0.589315239,2.48140259)\\
$E_{0}^{\xscrpt{N}^{4}\xscrpt{LO}}$ & (7.66667227,0.981322456)\\
\hline\hline
\end{tabular}
\end{center}
\caption[]{Coefficients of the perturbative expansion
(Eq.~(\ref{empole})) for the first radially excited pole in
$P$-wave $DD$ scattering.}
\label{DDcoefficients}
\end{table}

Figure~\ref{figDD}(b) shows that the pole position found
in the leading-order approximation agrees
with the full calculation only for very small values of $\lambda$,
but as the coupling increases,
the approximate pole starts to deviate strongly from the exact one
shown in Fig.~\ref{figDD}(a).
For example, at $\lambda$ = 0.5,
the first-order pole comes out below threshold but with a large imaginary
part, which is obviously unphysical. For the higher-order approximations,
the results are even worse, with the pole moving into the upper half plane,
or extremely deep down in the lower half for the N$^2$LO case. It becomes
evident that no perturbative approximation will produce anything even
resembling a bound-state pole for $\lambda\sim1$.

\subsection{$DK$ S-wave scattering}
\label{DKscattering}

\begin{figure}[htbp]
\centering
\includegraphics[scale=0.5]{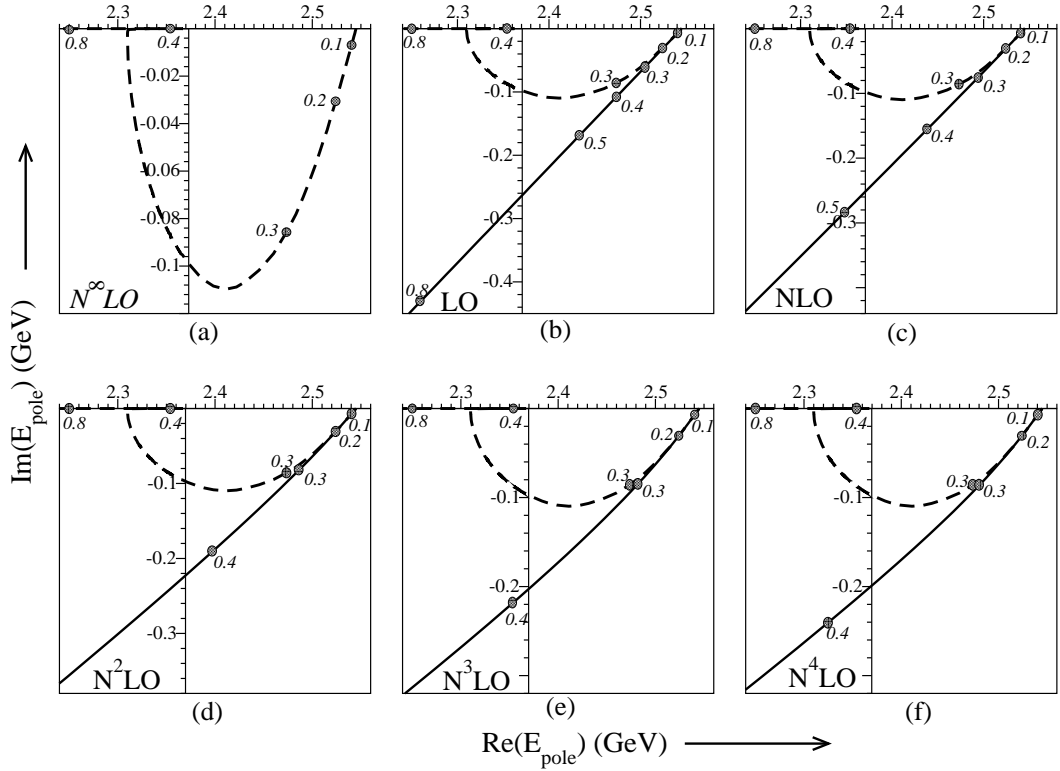}
\caption{\small
$S$-wave $DK$ pole positions in the complex-energy plane
(corresponding to the first and second Riemann sheets),
under variation of the coupling $\lambda$.
The dashed curves in all the plots for N$^{n}$LO ($n=0$, $\dots$, 4)
are the same as the one for N$^{\infty}$LO (labeled (a)).
The solid curves represent the perturbative results:
(b) leading (N$^{0}$LO, Born) term,
(c) next-to-leading (N$^{1}$LO),
(d) next-to-next-to-leading (N$^{2}$LO),
(e) (next-to)$^{3}$ (N$^{3}$LO),
and (f) (next-to)$^{4}$-leading (N$^{4}$LO) orders, respectively.}
\label{figDK}
\end{figure}

Finally, we study the case of $S$-wave $DK$ scattering
taking 
\begin{equation}
g_{n} = 2^{-n}\left( n+1 \right)^{1/2}.
\end{equation}
Now, as one can see in Fig.~\ref{figDK}(a),
the shape of the pole trajectory near the $DK$ threshold
is very different from the two $P$-wave cases.
For increasing $\lambda$, the pole approaches
the real-energy axis below threshold,
moves along the axis towards threshold as a virtual bound state,
and then becomes a bound state, moving finally to lower and lower
energies.
There is a one-to-one relation of this complex-energy pole trajectory
to the equivalent one in the complex-momentum plane. Thus, a
virtual bound state moving towards threshold corresponds
to a momentum pole moving upwards along the negative imaginary axis,
passing through the origin when the virtual bound state becomes a
true bound state, at threshold. In the present case, the pole is still
on the negative imaginary axis for $\lambda=0.4$,
but already on the positive one for $\lambda=0.5$.
This phenomenon, which happens exclusively for $S$-wave scattering,
as can be seen from the effective-range expansion,
is well described in Refs.~\cite{Taylor,LNP211p331}.
For $\lambda\approx0.6$, the bound-state pole reproduces the
$D_{s0}(2317)$ mass.

The coefficients of Eq.~(\ref{empole})
for the case of $S$-wave $DK$ scattering
are collected in Table~\ref{DKcoefficients}.
\begin{table}[htbp]
\begin{center}
\begin{tabular}{||l|r||}
\hline\hline & \\ [-5pt]
coefficient & value (GeV)\\ [5pt]
\hline & \\ [-5pt]
$E_{0}$ & (2.545,0.)\\
$E_{0}^{\xscrpt{N}^{0}\xscrpt{LO}}$ & (-0.445872986,-0.67333385)\\
$E_{0}^{\xscrpt{N}^{1}\xscrpt{LO}}$ & (-1.36316635,-1.84200144)\\
$E_{0}^{\xscrpt{N}^{2}\xscrpt{LO}}$ & (-9.95402765,-8.57593239)\\
$E_{0}^{\xscrpt{N}^{3}\xscrpt{LO}}$ & (-68.4326606,-43.6873369)\\
$E_{0}^{\xscrpt{N}^{4}\xscrpt{LO}}$ & (-299.284654,-138.032657)\\
\hline\hline
\end{tabular}
\end{center}
\caption[]{Coefficients of the perturbative expansion
(Eq.~(\ref{empole})) for the ground-state pole 
in $S$-wave $DK$ scattering.}
\label{DKcoefficients}
\end{table}
One sees at a glance that these coefficients
do not promise any kind of convergence.
Indeed, upon inspecting Figs.~\ref{figDK}(b--f),
one notices that only for small values of $\lambda$
the perturbative pole positions agree with the exact ones.
However, for $\lambda\ge 0.3$, the discrepancies grow rapidly, and no 
significant improvement is observed for higher orders of perturbation
theory.

\section{Summary and conclusions}

We have studied the discrepancies between perturbative estimates
for resonance pole positions and the exact ones, in the context of
a simple soluble model for hadronic decay of a meson.
In none of the considered cases satisfactory results were obtained
with the perturbative method, and not even any significant 
improvement was found for increasing orders of perturbation theory.
In particular, for bound states below the lowest strong-decay
threshold, no perturbative approximation produced anything like
a pole close to the real-energy axis. But also in the case of a
normal and not even very broad resonance, namely the $K^\ast(892)$,
the perturbative approach failed completely.

These results should be a warning for quark-model builders, because of
two reasons. First of all, the found large real mass shifts are, as a
consequence of analyticity, inseparably connected to the generation of
the physical hadronic widths, as demonstrated here for the $K^\ast(892)$,
but shown already many years ago for a variety of mesons \cite{PRD27p1527},
and confirmed in several more recent papers referred to above. Therefore,
any spectroscopic conclusions based on single-channel, ``quenched''
quark models should be taken with a great deal of caution. The second
reason is that even those quark models which pay some attention 
to strong decay, usually do this by employing perturbative methods.
The results presented here make it clear that a completely
non-perturbative treatment of hadronic resonances and bound states
is required for a realistic description.

\section*{Acknowledgments}

This work was supported in part by the {\it Funda\c{c}\~{a}o para a
Ci\^{e}ncia e a Tecnologia} \/of the {\it Minist\'{e}rio da Ci\^{e}ncia,
Tecnologia e Ensino Superior} \/of Portugal, under contract
CERN/\-FP/\-109307/\-2009.

\newcommand{\pubprt}[4]{{#1 {\bf #2}, #3 (#4)}}
\newcommand{\ertbid}[4]{[Erratum-ibid.~{#1 {\bf #2}, #3 (#4)}]}
\def\AIPCP{AIP Conf.\ Proc.}
\def\AP{Ann.\ Phys.}
\def\EPJA{Eur.\ Phys.\ J.\ A}
\def\EPJC{Eur.\ Phys.\ J.\ C}
\def\EPL{Europhys.\ Lett.}
\def\IJTPGTNO{Int.\ J.\ Theor.\ Phys.\ Group Theor.\ Nonlin.\ Opt.}
\def\LNP{Lect.\ Notes Phys.}
\def\NPA{Nucl.\ Phys.\ A}
\def\NPPS{Nucl.\ Phys.\ Proc.\ Suppl.}
\def\PLB{Phys.\ Lett.\ B}
\def\PRC{Phys.\ Rev.\ C}
\def\PRD{Phys.\ Rev.\ D}
\def\PRL{Phys.\ Rev.\ Lett.}
\def\ZPC{Z.\ Phys.\ C}

\end{document}